\documentclass[preprint,nocopyrightspace,numbers]{sigplanconf}

\usepackage{url}
\usepackage{amssymb}

\usepackage{gensymb}
\usepackage{subfigure}
\usepackage{flushend}
\usepackage[font=small]{caption}

\usepackage{color}
\definecolor{azure}{RGB}{51, 102, 153}
\definecolor{light}{RGB}{224, 224, 224}
\definecolor{lightblue}{RGB}{204, 204, 255}

\usepackage{tikz}
\usetikzlibrary{arrows,shadows,positioning,automata, calc, shapes.misc}

\usepackage{cleveref}

\makeatother

\begin{document}

\setlength{\pdfpageheight}{\paperheight}
\setlength{\pdfpagewidth}{\paperwidth}

\conferenceinfo{HLPC'16}{April 2, 2016, Atalanta, GA, USA}
\copyrightyear{2016}
\copyrightdata{978-1-nnnn-nnnn-n/yy/mm}
\copyrightdoi{nnnnnnn.nnnnnnn}

\titlebanner{~}  
\preprintfooter{~} 

\title{A Testbed for Transiently Powered Computers}

\authorinfo{Henko Aantjes \and Amjad Y. Majid \and Przemys{\l}aw Pawe{\l}czak}
{Delft University of Technology, Mekelweg\,4, 2628\,CD Delft, the Netherlands}
{Email: \{a.y.majid, p.pawelczak\}@tudelft.nl, h.aantjes@student.tudelft.nl}

\maketitle

\begin{abstract}
Transiently Powered Computers (TPCs) are novel devices that are battery-less and operate using only ambient energy. Therefore TPCs are prone to frequent power interruptions, and such the need for developing TPC-centric algorithms is a necessity. We advocate that only through a common experimental environment, accessible to everyone, a proper comparison of newly developed algorithms for TPCs can be realized. Moreover, only through access to various TPC testbeds---distributed geographically throughout the world---a proper applications testing and validation are possible. We enlist properties and features that any TPC testbed should have, calling for more coordinated action in this domain of TPC research. Finally, we present (to the best of our knowledge) world's first Internet-accessible testbed for TPCs.
\end{abstract}



\section{Introduction}
\label{sec:introduction}

Transiently Powered Computer (TPC) is an emerging wireless technology that aims at realizing battery-less portable computing devices. An example of TPC is Computational Radio Frequency Identification (CRFID). CRFID tags harvest energy from either the RF-signal~\cite{intel_wisp} or their ambient environments~\cite{parks_sigcomm_2014}. In addition, CRFIDs do not transmit signals, instead they backscatter incident (or ambient) signals to convey information. Consequently,  CRFID requires extremely low energy to communicate.

The advantage of the CRFID over classical RFID tags is obvious: CRFIDs have the ability to compute, sense and actuate. Furthermore, CRFIDs can be superior to traditional sensor nodes, since traditional sensor nodes need a battery to operate the energy-hungry transceiver circuit. The battery does not only increase the size and weight of the node, but also increases the cost of fabrication and maintenance. CRFIDs do not need this active transceiver circuit by exploiting backscatter communication, which allow them to operate in a low energy regime. Passive CRFIDs rely only on harvested energy from a Radio Frequency source. Unfortunately, this source does not transmit continually, which makes CRFIDs very susceptible to frequent energy interruptions. 

In summary, CRFIDs thanks to being battery-less can forever be embedded in a structure or a living organism but are very challenging to work with due to the transient nature of their activity.

\textbf{Motivation---Why Testbeds for TPCs are Needed:} There are many obstacles limiting the widespread adoption of TPCs. We list two of them. First, developing a new CRFID application requires having an access to relatively expensive hardware, i.e. an RFID reader in particular. Then, the CRFID itself is not easily accessible or it requires self-fabrication using open source data. Therefore, limited accessibility of TPCs makes spreading of this technology slower. Second, as research on TPCs is fragmented (and still sparse) newly developed protocols are often not verifiable, as there is no common set of experiments or environments where TPC algorithms should be tested. An obvious solution to these two challenges is the \emph{introduction of TPC testbeds}. Such testbeds would enable developers to directly test their algorithms and applications against different setup scenarios, e.g. different distances between CRFID tag(s) and reader(s) and different kinds of CRFID(s).

\textbf{Contribution:} To remove these two obstacles pertaining to TCP experimentation and to speed up the deployment of TPCs worldwide we have developed (to the best of our knowledge) the world's first remotely accessible \textit{testbed for TPCs}. Our TPC testbed is designed to explore various aspects of TPC ecosystems/TPC networks. In addition we list a set of requirements and features that any TPC testbed should have.

The remainder of the paper is organized as follows. Related work is discussed in Section~\ref{sec:related_work}, while Section~\ref{sec:features_testbed} introduces main features that any TPC testbed should have. Section \ref{sec:architecture} describes the architecture of our TPC testbed implementation. Example experiments, generated using our testbed, are described and discussed in Section~\ref{sec:example_experiment}. Finally, the paper is concluded in Section~\ref{sec:conclusions}.

\section{Related Work}
\label{sec:related_work}

TPCs are still in their early development stages and it is not surprising that there is no testbed publicly available thus far. Therefore, we will refer to related work from similar domains.

\textbf{Wireless Sensor Network:} Wireless Sensor Networks (WSN) are complex systems. They involve many nodes that interact with each other, where every communication link between two nodes is affected by many parameters. Hence, capturing all the variables in a theoretical model is a very challenging task. For this reason, many research groups have devoted great effort to develop commonly accessible WSN testbeds and platforms. For the interested reader we refer to~\cite{testbed_survey} which provides a comprehensive overview of the existing WSN testbeds.

\textbf{Radio Frequency Identification:} In contrast to the huge number of WSN testbeds, there is a very limited number of RFID-based testbeds. We are aware only of~\cite{dual_freq_rfid_testbed}, which explicitly mentions RFID testbed as a keyword, where the authors present the design and implementation of a reconfigurable RFID testbed that has support for two different frequencies. However, this testbed is basically an RFID reader, from which a user can modify a set of parameters. Moreover, this testbed is not accessible over the Internet. To summarize, testbeds are very common in WSN, while extremely limited for the traditional RFIDs and nonexistent for CRFIDs (TPCs).

\section{Features of Transiently Powered Computer Testbed}
\label{sec:features_testbed}

A testbed should be designed to fulfill the needs of as many users as possible. In other words, a testbed should allow to test the performance of any CRFID/TCP application in any environment without the need of physical access to CRFID devices. Therefore it should be highly flexible in terms of test conditions and have a large set of possible hardware resources. To make the most out of a CRFID/TPC testbed it should have the following features.

\begin{enumerate}
\item \textit{Internet Accessibility:} The testbed should be accessible by any user and from any location with an internet connection. The testbed should provide a fair scheduling policy as it is impossible to simultaneously grant any number of users access to it.

\item \textit{Reprogramming:} It should be possible for the users to execute and upload any firmware on the CRFID/TPC belonging to TPC testbed. Changing the application on the CRFID/TPC should be easy, such that a developer can make small changes in the application and test it again by uploading the firmware. Additionally, the reprogramming should be fast and not dependent on extra hardware, such as Flash Emulation Tools, which would increase the cost of the whole testbed (for the testbed deploying party).

\item \textit{Test Setup Control:} The testbed should allow to configure various test cases including: 

\begin{enumerate}
\item type of CRFIDs, e.g. WISP 4.0 or 5.1;
\item type of antenna and their number (at the reader side);
\item type of RFID reader (and its firmware);
\item movement pattern/speed between the reader antenna and CRFIDs/TCPs;
\item the angle of the tag antenna against the antenna of the reader;
\item density/number of tags in the testbed;
\item influence of objects, e.g. walls;
\item partial control of environmental conditions, e.g. scheduling experiments when interference at RFID frequency channels from other radio systems are minimal (for example during the night).
\end{enumerate}

\item \textit{Security:} Using the testbed should be reliable (with minimum downtime), private and secure. No measurement data may be stolen or influenced by evildoers. It should be difficult for someone to overtake the testbed system, nor should it be possible to damage the testbed itself.

\item \textit{Time and Energy Measurements:} Finally, it should be possible to measure energy consumption of the various CRFIDs while executing their tasks and to measure the timing of events like communication messages or GPIO changes.
\end{enumerate}

\subsection{Example Use of a CRFID Testbed}

If the TPC/CRFID testbed would have all the above features implemented, the following possibilities arise for experimentation:

\begin{enumerate}
\item \textit{Test New Communication Protocols:} The testbed could be used to develop new communication protocols like: hybrid ambient/classical backscatter communication, broadcasting for TPCs, and secure eavesdropping-prone communication.

\item \textit{Test New CRFID Applications:} Testbeds would allow to compare the operation of new unforeseen applications like CRFID cloud computing, CRFID cloud storage, CRFID security management application and most importantly testing operating systems for TPCs.
\end{enumerate}

\section{TPC Testbed: the Architecture}
\label{sec:architecture}
 
Given the fact that TPCs are still in the early development stage we have developed a testbed that implements only the essential requirements of a functional testbed. Our testbed enables, among others: remote accessibility, reprogramming and different environment setups. 

\begin{enumerate}
\item \textit{Internet Accessibility:} A user can access the TPC testbed via SSH connection at any time---provided that the testbed is not in use already---from any location with an internet connection. Users ought to create accounts, and such, any misuse is detectable and may lead to the user account or IP address blocking. However, fairness of use is not guaranteed.

\item \textit{Reprogramming:} The TPC testbed facilitates Wisent~\cite{wisent2016} to enable CRFIDs' firmwares/applications reprogramming. To make use of this feature the application must meet the following requirements:

\begin{itemize}
\item \textbf{Abortion:} The application must be able to receive the special Wisent ``go to bios'' command and abort its own execution. This implies that the application should eventually be able to receive EPC Gen2 messages.

\item \textbf{Compilation:} In order to be parsed correctly by the Wisent bootloader, modifications are needed in the linker file---the file which defines the memory space for the executable. If this modification is not performed then the bootloadable application will overwrite the Wisent boot loader program.
\end{itemize}

Violating the above requirements may disable the reprogramming capabilities or result in an inoperative/unresponsive TPC tag.

\item \textit{Test Setup Control:} The user is able to choose among three antennas. Each antenna represents a different operational setup. The multiple tag environment consists of six tags and two antennas. The user can either choose CRFIDs having different distances to the reader antenna or choose CRFIDs' setup where the CRFIDs have the same distance to the reader but with different antenna angle positions. In the single tag environment only one CRFID tag is available---a WISP 5.1 at a fixed distance. This setup will be described in detail in Section~\ref{sec:architecture}.

\item \textit{Security:} Thanks to SSH connection, which is configured for one user at the time, the testbed privacy is ensured. However, the user himself is responsible for deleting privacy-related log files. Unfortunately, the user still is able to make the testbed useless for others by programming the CRFIDs with malicious software, such that reprogramming is not possible anymore. This is due to the fact that CRFIDs cannot provide a failsafe wireless triggered reset option.

\item \textit{Time and Energy Measurements:} Both the reader and the host computer can provide UTC timestamps to events, which can be used for timing measurements. Energy consumption measurements of individual CRFIDs during application execution is not possible in our testbed.
\end{enumerate}

After introducing the features of our testbed, we now present its architecture in detail. The testbed consists of two main parts: (i) the user side, (ii) and the server side. The user side is everything concerning the user connecting to the testbed. The server side is the physical hardware at which the tests can be performed.

\tikzset{cross/.style={cross out, draw=black, minimum size=2*(#1-\pgflinewidth),inner sep=0pt, outer sep=0pt},cross/.default={3pt}}
\tikzstyle{a} = [rectangle, draw, node distance=8mm, fill=white!20, text width=12mm, text centered, rounded corners, minimum height=2em, thick]
\tikzstyle{l} = [draw, -latex',thick]
\begin{figure}
	\centering
	\subfigure[Testbed hardware]{
		\begin{tikzpicture}[-,>=stealth', semithick, node distance = 15mm, scale=0.87, every node/.style = {scale=0.87}]
			\scriptsize
			
			\node [a] (A)   {Host (PC)};
			\node [a] (B) at ([shift={(0,-1)}] A) {Reader};
			\node [a] (C) at ([shift={(0,-1)}] B) {Tag};
			\node [a] (D) at ([shift={(0,1)}] A) {Server};
			\node [a] (E) at ([shift={(0,2)}] A) {User};
			
			\draw [draw=black,thin] ([shift={(0,0.12)}]B.east) -- ([shift={(0.3,0.12)}]B.east) |- ([shift={(0.3,0.7)}]B.east);
			\draw [draw=black,thin] ([shift={(0.3,0.4)}]B.east) -- ([shift={(0.4,0.7)}]B.east);
			\draw [draw=black,thin] ([shift={(0.3,0.4)}]B.east) -- ([shift={(0.2,0.7)}]B.east);
			
			\draw [draw=black,thin] ([shift={(0,0)}] C.west) -- ([shift={(-0.3,0)}]C.west) |- ([shift={(-0.3,0.7)}]C.west);
			\draw [draw=black,thin] ([shift={(-0.3,0.4)}]C.west) -- ([shift={(-0.4,0.7)}] C.west);
			\draw [draw=black,thin] ([shift={(-0.3,0.4)}]C.west) -- ([shift={(-0.2,0.7)}] C.west);
			
			\draw [draw=black] (A.south) -- (B.north);
			\draw [draw=black] (A) -- (D);
			\draw [draw=black] (D) -- (E);
			\draw [<->, draw=black,thin] (A.west) -- ([shift={(-0.3,0)}]A.west) |- ([shift={(-0.3,0)}]B.west) -- (B.west);
			\draw [<->, draw=black,thin] (B.east) -- ([shift={(0.3,0)}]B.east) |- ([shift={(0.3,0)}]C.east) -- (C.east);
			\draw [<->, draw=black,thin] (E.west) -- ([shift={(-0.3,0)}]E.west) |- ([shift={(-0.3,0)}]D.west) -- (D.west);
			\draw [<->, draw=black,thin] (A.east) -- ([shift={(0.3,0)}]A.east) |- ([shift={(0.3,0)}]D.east) -- (D.east);
			
			\node [thick, rotate=90] at (-1,-0.45) [above] {LLRP};	
			\node [thick, rotate=-90] at (1,-1.5) [above] {EPC C1G2};
			\node [thick, rotate=-90] at (1,0.5) [above] {SSH};
			\node [thick, rotate=90] at (-1,1.5) [above] {SSH};
		\end{tikzpicture}
	}
	\hspace{7mm}
	\subfigure[Photograph of a multiple tag environment]{\includegraphics[width=0.45\columnwidth,trim={0 0 0 23cm},clip]{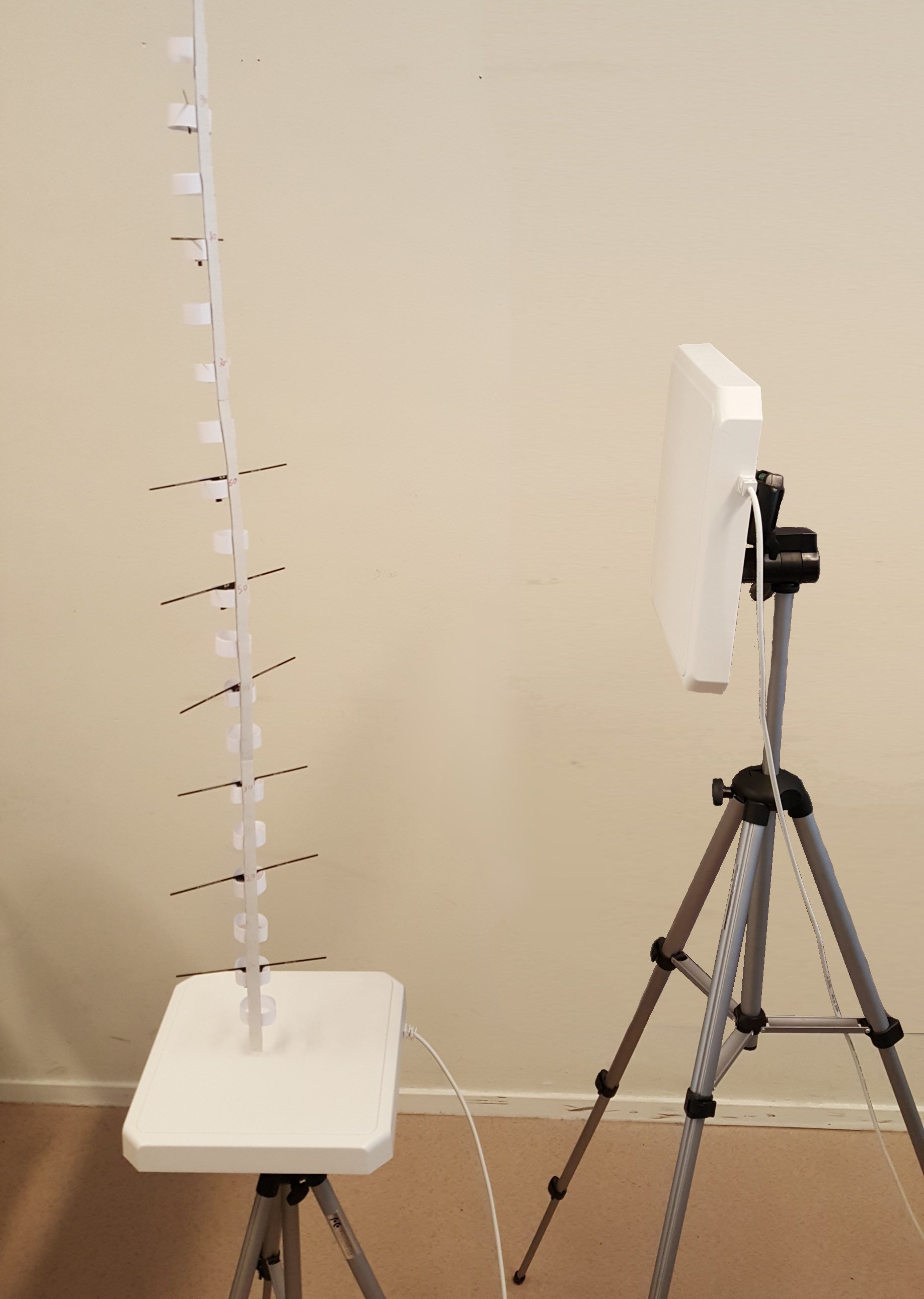}}
\nocaptionrule \caption{Testbed setup: (a) protocol layers of the TPC testbed (b) multiple tag environment setup. Antenna 2 (on the bottom) sees the tags at different distances, but with equal angle. For antenna 3 (on the right) all tags are at a fixed distance, but have different angles.}
\label{fig:arch}
\end{figure}
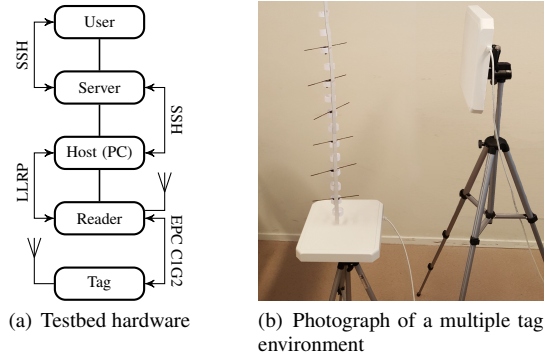

\subsection{Architecture: the Server Side}

Three fundamental components of the testbed system are (i) a host machine---any device that is able to communicate with the RFID reader over any popular physical interfaces such as Ethernet or Wi-Fi, (ii) an RFID reader (connected to one or more antennas) and (iii) one or more (C)RFID tags, see Fig.~\ref{fig:arch}.

\subsubsection{Communication Layers}

The host machine of our testbed is a Linux PC running Ubuntu 14.04 connected via the ethernet port to the RFID reader. The host machine translates the user commands via a Python script that communicates with the Reader. This communication between the Reader and the Host is over Ethernet, conforms to the Low Level Reader Protocol (LLRP)~\cite{llrp2010}, which is implemented in the SLLURP Python library~\cite{sllurp}. With the LLRP protocol the host can give tasks to the Reader, such as ``Inventory which tags are in the communication range of a specific antenna and report this back to the host after ten seconds.'' The reader executes the tasks by communicating with the tags according to the EPC C1G2~\cite{epc2015} standard.

\subsubsection{Physical Setup}

We use an Impinj R1000 RFID reader (Octane 3.2.4.240 firmware version) and connect it to three circularly-polarized antennas (CushCraft S9028PCRJ 8\,dBic). Each antenna is placed at a different position relative to the CRFID tags. We characterize the distance between antenna $x$ and the center of the CRFID tag $y$ as $l_{x,y}$, while the rotation of the CRFID tag $y$ from zero polarization against the antenna $x$ as $\theta_{x,y}$. Using these two parameters we provide testbed setup for each of three antennas.

\begin{itemize}
\item \textbf{Antenna 1} is used for a single RFID tag measurement consisting of one WISP 5.1 at an angle $\theta_{1,1} = 0\degree$ and a distance $l_{1,1}=20$\,cm;

\item \textbf{Antenna 2} is used in a multiple RFID tags environment consisting of 6 CRFIDs (again, WISP 5.1s), all at different distances to the reader but properly and equally aligned, i.e. $\theta_{2,y} = 0, \forall y$. The distance between the reader antenna $x=2$ and a tag $y$ is $l_{2,y} = 10(y+1)$\,cm with $y=[0,5]\in\mathcal{N}$;

\item \textbf{Antenna 3} is directed towards the same multiple tags environment as Antenna 2, but such that all tags have a different angle to the reader, while having the same distance to the reader, i.e. $l_{2,y} = 30$ cm$, \forall y$. The angle for tag $y$ is $\theta_{y,3} \approx 10\degree y$ with $y=[0,5]\in\mathcal{N}$.
\end{itemize}

Finally, all antennas are placed  70\,cm away from a wall and the testbed has no other objects in a $\approx$1.5\,m range. This physical setup makes it possible to select the environment by selecting one of the antennas. Because of the various CRFIDs at various distances or various angles in the multiple RFID tags environment it is possible to test an application under all these different conditions by selecting the corresponding antenna.

\subsection{Architecture: the Client Side}


A testbed user has several options\footnote{To obtain the SSH access to the testbed please contact the authors of this paper.}. He may change the testbed setup, i.e. choosing between the multiple and singular operational modes. The user may reprogram the CRFID's with a new application, by the use of the Wisent protocol. After performing these steps the user can execute one of the default Python scripts or his own Python script to communicate with the tags and test the (new) application. While executing the program the user receives instant log messages on the performance. The timing information will be provided by the Python script with a timestamped log file. The UTC timestamps have a millisecond resolution.

\section{TPC Testbed: an Example Experiment}
\label{sec:example_experiment}

To show the possibilities of our testbed we conducted two example experiments described below: (i) inventory experiment, and (ii) CRFID reprogramming. 

\subsection{Inventory Experiment}
\label{sec:inventory_experiment}

\begin{figure}
	\centering
	\includegraphics[width=\columnwidth]{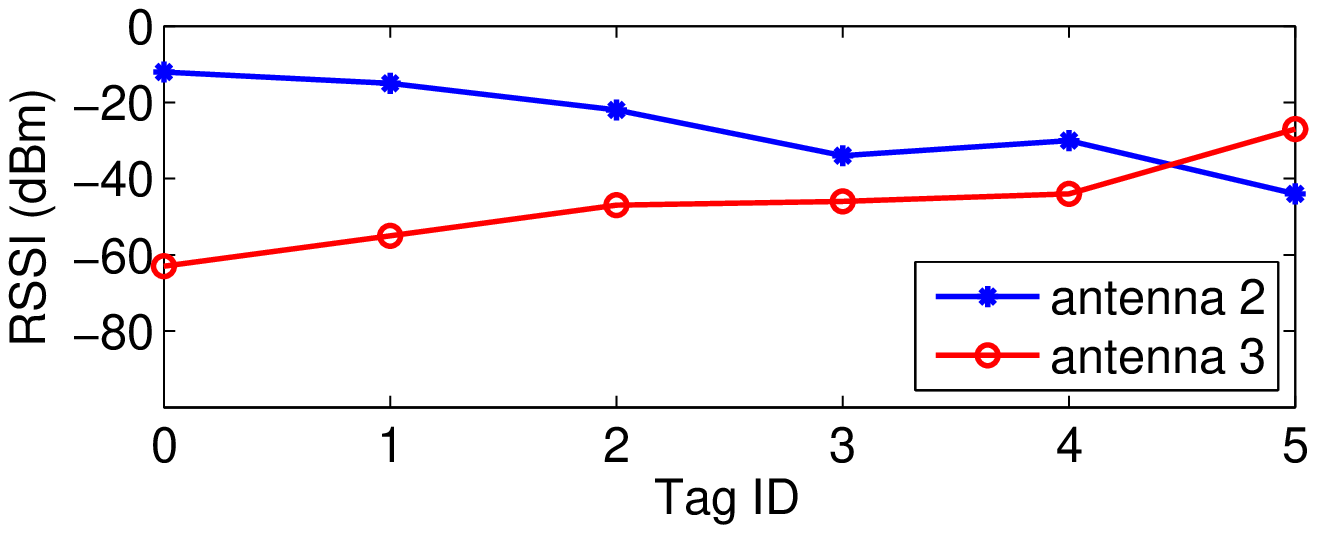}
	\includegraphics[width=\columnwidth]{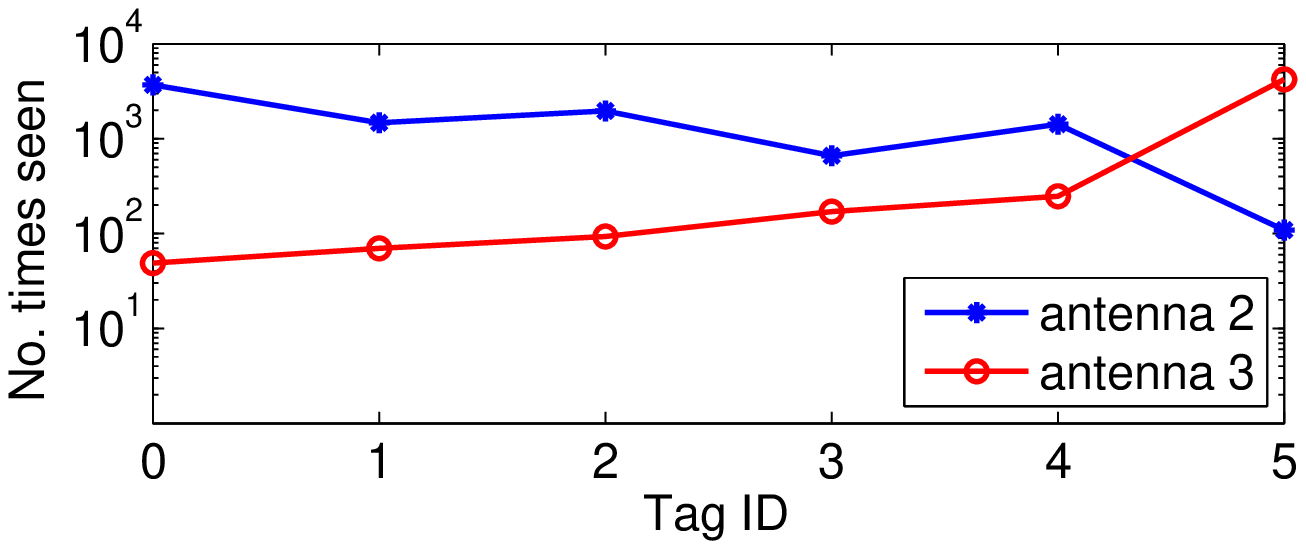}
	\nocaptionrule \caption{Testbed inventory results for the two antennas in the multiple tag environment: (top) mean RSSI values for each tag, (bottom) number of times the tag communicated during the 30 seconds experiment. }
	\label{fig:inventory}
\end{figure}

For the user it might be interesting to know which tags are in the communication range of the reader and what the quality of this communication link is. Therefore we added a simple inventory script to the testbed, which shows in real-time which tags are in the inventory of the testbed and via which antenna the tag is reachable. Moreover it provides for each tag-antenna combination the communication quality in terms strength and aliveness of the channel, respectively the mean RSSI value and the number of times that the tag communicated during the experiment. The inventory script is run for half a minute for each antenna. The result of this experiment can be found in Section~\ref{fig:inventory}.

As can be seen in the Figure, for antenna 2 holds that the RSSI values become lower for the tags that are further away, which also results in a lower number of connections. However the RSSI values to antenna 3 cannot be explained solely by considering the angles. The low RSSI value and lower number of communications for the tags with lower ID is expected, because they have an angle close to 90 degrees. The communication antenna of the WISP is a dipole, which means that communication at 90 degrees is theoretically impossible. The extreme high number of times that tag 5 is seen compared to the slightly rotated other tags is unexpected. What most likely caused this effect is the environment with multitude of tags. The tags in the middle will have the highest influence from other tags and tag 5 benefits from less interference from surrounding tags.

\subsection{CRFID Reprogramming}
\label{sec:reprogramming}

One crucial feature of a testbed is the possibility to load an application onto a CRFID without physical intervention of a Flash Emulator Tool. For this we have used Wisent protocol that enables such functionality and integrated it in the testbed. Off course it is important to know how long it takes to reprogram the CRFIDs in the testbed, so we tested this by reprogramming each CRFID separately with a basic application. To increase the reprogramming speed we used the antenna which has the best communication link. For tags 3 and 4 the highest speed was established by enabling both the antennas 2 and 3. 

The average time to reprogram a single CRFID with antenna 1 is 105 seconds. Whereas, 
the time to reprogram a set of CRFIDs is given in Section~\ref{wrep_times}. The significant reprogramming time suggests that further improvements for Wisent are required.
One slightly surprising result is the long reprogramming time of tag 0 although it was within 20\,cm from the antenna. This could be explained by the fact that the WISP is designed to operate in the far field operational region, which is above 30\,cm for the UHF band. 

\begin{table}[t!]
\centering
\caption{Reprogramming times for a set of CRFID tags in the testbed; Refer to Section~\ref{sec:reprogramming} for details.}
\begin{tabular}{| c | c | c |}
\hline
WISP no. & Antenna & Reprogramming time [s] \\ \hline\hline
0 & 2 & 1 343 \\ \hline
1 & 2 & 112 \\ \hline
2 & 2 & 271 \\ \hline
3 & 2 and 3 & 1 614 \\ \hline
4 & 2 and 3 & 2 168 \\ \hline
5 & 3 & 628 \\ \hline
\end{tabular}
\label{wrep_times}
\end{table}

\section{Conclusions}
\label{sec:conclusions}

In this paper we have advocated for a creation of Transiently Powered Computer (TPC) testbed. We have listed fundamental features of such a testbed and speculated what experiments can be executed. Then we have presented the first implementation of such testbed. Naturally, we are aware that the current implementation of the TPC testbed lacks many features that are required. Nevertheless we are confident that this is a first step towards a realization of a fully-functional and extensively used remotely accessible TPC testbed. Our TPC testbed enables the developers to completely reprogram a TPC tag, i.e. WISP. Additionally, a user can choose between a multiple or a single tag setup. Finally, a user can obtain detailed logging information, for example timing, RSSI values and how many times a tag has been seen in a single experiment. 



\bibliographystyle{plainnat}
\bibliography{IEEEabrv,references}

\end{document}